\documentclass[twocolumn,prl,aps,epsf,showpacs]{revtex4}

\usepackage{amssymb,epsf}
\usepackage{graphicx}
\usepackage{epstopdf}

\begin{document}


\author{C.R. Dean$^{1}$, B.A. Piot$^{1}$, P. Hayden$^{2}$, S. Das Sarma$^{3}$, G. Gervais$^{1}$, L.N. Pfeiffer$^{4}$ and K.W. West$^{4}$ }

\address{$^{1}$Department of Physics, McGill University, Montreal, H3A 2T8, CANADA}
\address{$^{2}$School of Computer Science, McGill University, Montreal, H3A 2A7, CANADA}
\address{$^{3}$Condensed Matter Theory Center, Department of Physics, University of Maryland, College Park, MD 20742 USA}
\address{$^{4}$Bell Laboratories, Alcatel-Lucent Inc., Murray Hill, NJ 07974 USA}



\title{Contrasting  Behavior of the $\frac{5}{2}$ and $\frac{7}{3}$ Fractional Quantum Hall Effect in a Tilted Field}

\begin{abstract}

Using a tilted field geometry, the effect of an in-plane magnetic field on the even denominator $\nu=\frac{5}{2}$ fractional quantum Hall state is studied. The energy gap of the $\nu=\frac{5}{2}$ state is found to collapse linearly with the in-plane magnetic field above $\sim 0.5$T. In contrast, a strong enhancement of the gap is observed for the $\nu=\frac{7}{3}$ state.  The radically distinct tilted-field behaviour between the two states is discussed in terms of Zeeman and  magneto-orbital coupling within the context  of the proposed Moore-Read Pfaffian wavefunction  for  the $\frac{5}{2}$ fractional quantum Hall effect.

\end{abstract}
  \maketitle

The incompressible even-denominator fractional quantum Hall effect (FQHE) at filling factor $\nu=\frac{5}{2}$  \cite{Willett:PRL:1987} remains one of the most exotic phenomena ever discovered in condensed matter physics.  Of the $\sim$100 FQH states observed in high-mobility  two-dimensional electron gas (2DEG) structures, the $\frac{5}{2}$  FQH state continues to represent the only violation of the odd-denominator rule. Aside from the fundamental questions this even-denominator FQH state raises concerning our understanding of strongly correlated electron behaviour, the $\frac{5}{2}$ state has received particular attention because of its possible description by the  Moore-Read Pfaffian wave function \cite{MooreRead:NuclPhysB:1991}, which exhibits non-abelian quantum statistics.  This potential non-abelian property makes the $\frac{5}{2}$ FQHE a strong candidate for the realization of fault-tolerant topological quantum computation \cite{Nayak:PRL:2005}. Recent measurements of the $e^{*}=e/4$ quasi-particle charge \cite{Dolev:Nat:2008}, as well as  the tunneling spectra \cite{Radu:arXiv:2008}, are fully consistent with a Moore-Read Pfaffian wavefunction (or its particle-hole conjugate, the so-called anti-Pfaffian \cite{Levin:PRL:2007, Lee:PRL:2007}). However, an unequivocal experimental proof is still lacking, in part due to other possible candidate paired-states, such as the (abelian) Halperin (3,3,1) state, which could {\it in principle}
be realized at  $\nu=\frac{5}{2}$. Understanding the nature of the $\frac{5}{2}$ FQHE thus remains an important open question \cite{Jain2006,Jain2007}.

Key features of the Moore-Read Pfaffian that distinguish it from other possible wavefunctions include its fully spin-polarized electron polarization \cite{Dimov:PRL:2008, Feiguin:arXiv:2008,Morf:PRL:1998}, as well as the non-zero angular momentum of its $p_{x}+\imath p_{y}$ pairing, very similar to the $A_{1}$-phase of superfluid $^{3}$He. In this Letter, we investigate the $\frac{5}{2}$ FQH  spin polarization by measuring the activation energy gap of the  $\frac{5}{2}$ (and particle-hole
conjugate $\frac{7}{2}$) and neighbouring $\frac{7}{3}$ state in a tilted field geometry where, in addition to the perpendicular field (B$_{\perp}$), a parallel field (B$_{\parallel}$) is applied in the plane of the 2DEG. Two important aspects distinguish our work from previous tilted field experiments in the second Landau level (SLL) \cite{Eisenstein:PRL:1988, Eisenstein:SurfSci:1990, Lilly:PRL:1999, Pan:PRL:83:1999, Csathy:PRL:2005}.  First, we examine a relatively strong $\frac{5}{2}$-state occurring at lower  magnetic field than previously observed (owing to our low density sample \cite{Dean:PRL:2008}) in the non-perturbative Landau level coupling regime where the Coulomb interaction energy is greater than the Landau level separation.
 Secondly, we study the FQH energy gap in a sample with a relatively wide 2D quantum well so that under tilt the magnetic length associated with B$_{\parallel}$  becomes of order or smaller than the well width, allowing us to study the effect of  orbital coupling to the parallel field.  Comparing the even-denominator $\frac{5}{2}$ with the odd-denominator $\frac{7}{3}$, we find that the in-plane magnetic field induces dramatically (and qualitatively) different behaviour in the two activation gaps.  This is a particularly surprising result since current theoretical models that interpret the decreasing $\frac{5}{2}$ gap under tilt in the context of orbital coupling to the parallel field, suggest the same gap suppression should be seen in the neighbouring $\frac{7}{3}$ state \cite{Morf:PRL:1998, Peterson:arXiv:2008}.  The unexpected experimental finding presented here, in direct contradiction to theory, implies that our understanding of the FQHE in the second landau level is incomplete

Our experiment was conducted in a  40~nm wide, modulation-doped, GaAs/AlGaAs quantum well, with a measured density of 1.6(1)$\times$10$^{11}$~cm$^{-2}$ and mobility of 14(2)$\times$10$^{6}$~cm$^{2}$/V·s. The sample was cooled in a Janis  JDR-100 dilution refrigerator equipped with a 9 Tesla magnet. Treatment with a red LED was used during the cooldown. Details of the sample cooling and temperature measurement have been described elsewhere \cite{Dean:PRL:2008}.  The tilted field geometry was achieved by rotating the sample \textit{in situ} in a static field.
All transport measurements were performed using a standard lock-in technique at $\sim$6.5~Hz and small excitation current, I$_{exc}$ = 2 - 10~nA.

\begin{figure}[t]
	\begin{center}
		\includegraphics[width=0.7\linewidth, angle=0,clip]{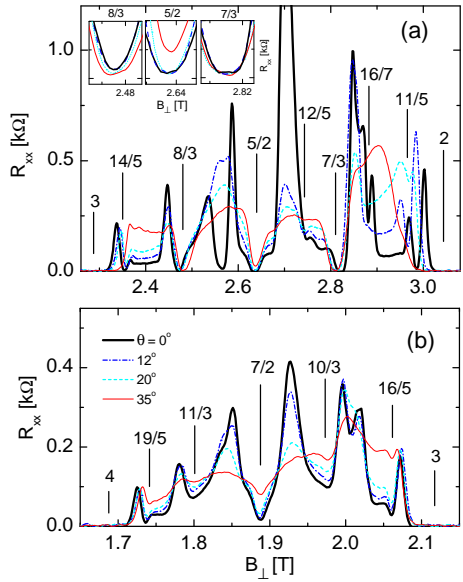}
		\caption{Magnetoresistance under tilt in the second Landau level around (a) $\nu=\frac{5}{2}$ and (b) $\nu=\frac{7}{2}$ plotted versus perpendicular field (T $\sim$ 20~mK). Inset:  enlargement showing the trend of the $\nu=\frac{8}{3}$, $\frac{5}{2}$, and $\frac{7}{3}$ FQHE minima.}
		\label{Fig1}
	\end{center}
\end{figure}

\begin{figure}[th]
	\begin{center}
		\includegraphics[width=1\linewidth,angle=0,clip]{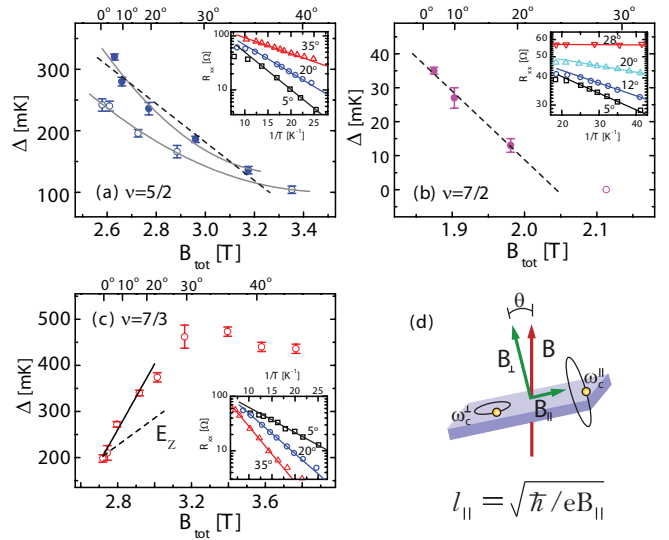}
		\caption{(a)-(c) Activation gaps under tilt (corresponding Arrhenius plots shown in inset).  Two data sets are shown in (a) with the open symbols corresponding to a slightly smaller density and mobility than the solid symbols.  Open symbol in (b) was taken from an already destroyed gap.  In (a), (b) linear fits (dashed lines) are used to estimate the g-factor (see text); solid curves are a guide to the eye.  In (c ) solid line is a linear fit to the low B-field data; dashed line is for a single-particle Zeeman interaction. (d) Schematic of a 2DEG in a tilted field.}
		\label{Fig2}
	\end{center}
\end{figure}

Fig. \ref{Fig1} shows the magnetoresistance in the SLL at various tilt angles, with $\theta$ referring to the angle between the applied field and the normal of the 2DEG plane.  The behaviour of the $\frac{5}{2}$ minima  as a function of 
the tilt angle $\theta$  is strikingly different from  the neighbouring, odd-denominator, $\frac{7}{3}$ and $\frac{8}{3}$ states. As emphasized in the inset of Fig. \ref{Fig1}a, the $\frac{5}{2}$ minimum clearly diminishes while the   $\frac{7}{3}$ and $\frac{8}{3}$ strengthen with tilting. The suppression of the $\frac{5}{2}$ state by a parallel field has been known since the earliest tilted field experiments \cite{Eisenstein:PRL:1988, Eisenstein:SurfSci:1990}, which was shown later on to yield to a stripe phase at high tilt
 \cite{Lilly:PRL:1999,Pan:PRL:83:1999}. However, while several theoretical treatments have proposed this 
to be related to magneto-orbital  B$_{\parallel}$-induced destruction of the $\frac{5}{2}$ FQHE\cite{Morf:PRL:1998, Rezayi:PRL:2000, Peterson:arXiv:2008},  little effort has been made to examine whether or not the same effect is manifest in the neighbouring odd-denominator states \cite{Eisenstein:PRL:1988, Eisenstein:SurfSci:1990, Lilly:PRL:1999, Pan:PRL:83:1999, Csathy:PRL:2005}.  Where qualitative observations have been made concerning the $\frac{7}{3}$, they appear inconsistent with, for example, the state reported to strengthen \cite{Eisenstein:PRL:1988}, weaken \cite{Morf:PRL:1998,Eisenstein:SurfSci:1990}, and/or `stay robust' \cite{Csathy:PRL:2005} under tilt. 

We measured the activation energy gap as a function of tilt angle for the $\frac{5}{2}$, $\frac{7}{2}$ and $\frac{7}{3}$ states.  The energy gap, $\Delta$,  was determined from the temperature dependence of the corresponding resistance minima in the thermally activated regime where $R_{xx}\propto e^{-\Delta/2k_{B}T}$ .  In Fig. \ref{Fig2} the gaps are plotted versus total magnetic field, $B_{tot}$,  with a set of temperature dependence curves (each data point was acquired at fixed field) shown in an Arrhenius plot inset in each panel.  Consistent with the qualitative trend seen in Fig. \ref{Fig1}, the $\frac{5}{2}$ and $\frac{7}{2}$ gap diminishes with tilt while the $\frac{7}{3}$ shows a remarkable enhancement, producing a gap larger by more than a factor of two at only $\theta\sim$35$^{\circ}$.  
It is important to note that in contrast to refs \cite{Lilly:PRL:1999, Pan:PRL:83:1999} varying the measurement configuration in our sample did not show any evidence of anisotropy up to the highest tilt angle investigated ($\theta=44^{o}$).

In a perfectly 2D system, {\it i.e.} with a vanishing quantum well width, applying an in-plane field by sample rotation should couple only to the electron spin through the Zeeman energy, E$_{Z}$=g$^{*}$$\mu_{B}$B$_{tot}$$\cdot$S, where g$^{*}$ is the effective electron g-factor, $\mu_{B}$ the Bohr magneton, S the electron spin and B$_{tot}$ is the \textit{total} applied field. As such, the most obvious interpretation of our observed weakening $\frac{5}{2}$ (and $\frac{7}{2}$) states compared with the strengthening $\frac{7}{3}$ (and $\frac{8}{3}$) FQHE is that  the $\frac{1}{3}$ SLL FQH states ($\frac{7}{3}$ or $\frac{8}{3}$) are spin-polarized whereas
the $\frac{5}{2}$ and $\frac{7}{2}$ states are spin-unpolarized. 
A spin-unpolarized $\frac{5}{2}$ state however would be in conflict with extensive theoretical evidence in favor of  a fully spin-polarized state, most likely described by the Moore-Read Pfaffian wave function \cite{Morf:PRL:1998,Dimov:PRL:2008, Feiguin:arXiv:2008}. Moreover, in a single-particle Zeeman picture, a reduction of the $\frac{5}{2}$ gap induced by spin coupling alone should yield a universal slope in a plot of activation gap versus $B_{tot}$, determined by the effective g-factor whose commonly accepted value in GaAs is $g^{*}=-0.44$. Extracting the g-factor from our $\frac{5}{2}$ and $\frac{7}{2}$ gaps, we obtain values  of 0.47$\pm$0.06 and 0.31$\pm$0.03 respectively, compared with the value 0.56 reported in the work
of Eisenstein {\it et al.}\cite{Eisenstein:SurfSci:1990}. Additionally, closer inspection of the $\frac{5}{2}$ data (Fig. 2a) suggests a linear trend (dotted line) does not fit the data well over the entire range, but rather the data is non-linear (solid curve guide to the eye),  raising further doubts about the spin-induced gap suppression in a spin-unpolarized FQHE interpretation. In our view, the non-linear behaviour in $B_{tot}$, together with the scattered g-factors, is indicative of a gap suppression not driven by Zeeman coupling, but rather  by another mechanism. In a real 2DEG sample with finite thickness, orbital coupling between the in-plane field and the transverse electron dynamics becomes possible when the parallel field magnetic length becomes of order or less than the quantum well width. In this case, a spin-polarized $\frac{5}{2}$ state can also be suppressed with increasing B$_{\parallel}$ \cite{Morf:PRL:1998, Rezayi:PRL:2000, Peterson:arXiv:2008} since the activation gap is no longer determined by Zeeman coupling alone. At B$_\parallel$=1~T, the parallel magnetic length  $l_{\parallel}$=$\sqrt{\hbar/{eB}_{\parallel}}$ is only 26 nm (already nearly half the width of our quantum well) making it plausible for magneto-orbital effects to play an important role in our experiment, even under modest tilting.

While the magneto-orbital coupling for $B_{\parallel}\neq 0$ may explain \cite{Morf:PRL:1998, Rezayi:PRL:2000, Peterson:arXiv:2008} the suppression of the $\frac{5}{2}$ FQHE, it should also  destabilize the $\frac{7}{3}$ in a similar way. In his seminal work where Morf proposed an explanation for the  $\frac{5}{2}$ gap suppression driven by orbital coupling to the parallel field  \cite{Morf:PRL:1998} he argued that a similar suppression would be expected for the neighbouring $\frac{7}{3}$ state.  In a numerical study, Peterson \textit{et al.} recently reported the $\frac{7}{3}$ being strengthened 
by the increase of the finite thickness of the 2DEG \cite{Peterson:arXiv:2008}. In agreement with Morf, they therefore concluded that coupling to a parallel field, whose main effect is well known to squeeze the wavefunction towards the ideal 2D limit, is expected to weaken the $\frac{7}{3}$ state similar to the $\frac{5}{2}$.  In both of these studies a Pfaffian was used for the $\frac{5}{2}$ wavefunction whereas a Laughlin wavefunction was taken for the $\frac{7}{3}$.  The theoretical predictions therefore follow despite the two states being described by radically different ground states.
All of this theoretical work is in disagreement with the unambiguous enhancement of the $\frac{7}{3}$ gap observed in our work.  We note that the gap increase in  Fig. \ref{Fig2}c is non-linear, with the low-tilt regime exhibiting a slope (solid line) nearly $\sim$2.5 times \textit{larger} than expected by single-particle Zeeman coupling alone (dashed line). The growth of the $\frac{7}{3}$ gap at a greater rate than predicted by single-particle Zeeman interaction alone suggests a non-trivial excitation spectrum, perhaps involving a spin texture similar to the skyrmion excitations observed at $\nu=\frac{1}{3}$ in the lowest landau level \cite{Leadley:PRL:1997, Dethlefsen:PRB:2006}. The formation of skyrmions is characterized by the interplay between Zeeman and Coulomb energies, with skrymionic spin-reversal excitations existing only below some critical value of the ratio $\eta=\frac{E_{Z}}{E_{C}}$.  In our sample $\eta\sim0.01-0.013$ at $\nu = \frac{7}{3}$, well below the critical value of $\eta_{c}=0.022$ at $\nu=1$ \cite{Schmeller:PRL:1995} and similar to the value of $\eta_{c}=0.01$ measured at $\nu=\frac{1}{3}$ \cite{Leadley:PRL:1997}.  It is therefore possible that our observed $\frac{7}{3}$ behavior arises from the existence of small skyrmions involving two to three reversed spins.

\begin{figure}[t]
	\begin{center}
		\includegraphics[width=0.8\linewidth,angle=0,clip]{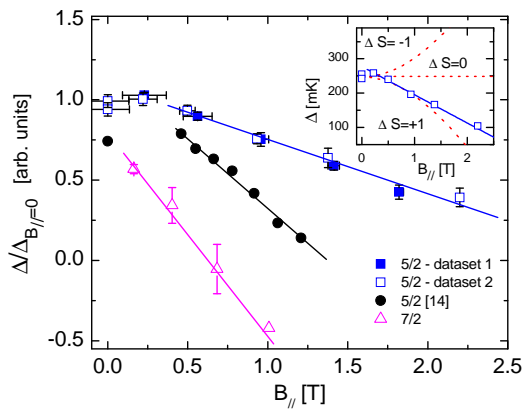}
		\caption{Normalized gap values versus parallel field (datasets are vertically offset for clarity).  All data is replotted from Fig. 2 except the solid circles which is replotted from Ref.\cite{Eisenstein:SurfSci:1990}.  Inset:  Experimental gap versus B$_{\parallel}$ at $\nu=\frac{5}{2}$  (open squares) compared with the theoretical interaction for single spin flips (dashed lines, see text).}
		\label{Fig3}
	\end{center}
\end{figure}

In Fig. \ref{Fig3} the $\frac{5}{2}$ gap values from Fig. 2 are replotted versus the parallel field, $B_{\parallel}$, and normalized
to their zero-tilt values.  Whereas the $\frac{5}{2 }$ gaps appear to vary non-linearly with changing B$_{tot}$ (Fig \ref{Fig2}), they follow a remarkably linear trend versus B$_{\parallel}$.  We also note in the $\frac{5}{2}$ data (open and closed squares) a range in low B$_{\parallel}$ distinct from the high B$_{\parallel}$ where the gap is unmodified for tilt up to $\sim 10^{\circ}$.  Interestingly, a similar trend is also observed when replotting the data from reference \cite{Eisenstein:SurfSci:1990} (closed circles). The onset of the gap suppression for the $\frac{5}{2}$ state occurs in all data set at 
a field strength  B$_{\parallel}$$\sim$0.5~T, corresponding to a transverse magnetic length $l_{\parallel}$=36~nm. This is remarkably close to the width of our quantum well and is therefore consistent with magneto-orbital coupling  effects becoming relevant when the transverse magnetic length approaches the size of the quantum well. Importantly, a constant gap value is not observed in the $\frac{7}{3}$ data at low tilt.  We take this as a further indication that the $\frac{5}{2}$ gap suppression and the $\frac{7}{3}$ gap enhancement originate from distinct mechanisms.  
We show in the inset of Fig. \ref{Fig3} our measured $\frac{5}{2}$ gap data versus $B_{\parallel}$ against the theoretical trend expected for a Zeeman energy variation associated with a total spin change $\Delta S = -1,0$  (dashed increasing, flat line), and $\Delta S= +1$ (decreasing dashed line), expected for a spin-polarized and spin-unpolarized ground state, respectively. While, within the resolution of our experiment,  we cannot distinguish whether the low B$_{\parallel}$ data better fits the trend for a spin-polarized  or unpolarized ground state, 
the gap suppression of the $\frac{5}{2}$ state following a linear behaviour in $B_{\parallel}$, and departing
from the $\Delta S=+1$  curve at field $B_{\parallel}\gtrsim 1.5$T, is further evidence for a gap suppression driven by magneto-orbital rather than Zeeman coupling effects.

A theoretical treatment of the complex electron dynamics in the presence of the strong in-plane field \cite{Stopa:PRB:1989} is well beyond the scope of our experimental work.  However, with the $\frac{5}{2}$ and $\frac{7}{3}$ states appearing so closely together in magnetic field, having  nearly the same energetics and Landau level coupling when B$_{\parallel}$=0, we speculate that the contrast in behaviour between the $\frac{5}{2}$ and $\frac{7}{3}$ states may require some fundamentally new theoretical insight. One possibility is that this difference arises, not from a ground state spin-polarization difference between these two states, but rather from a difference in the excited states. If we consider both the $\frac{5}{2}$ and $\frac{7}{3}$ states to  be spin polarized, then the decreasing $\frac{5}{2}$ gap compared to the increasing $\frac{7}{3}$ gap could be explained by the $\frac{7}{3}$ trend resulting from a combination of spin-reversed excitation (gap enhancement) and finite width (gap suppression) whereas the $\frac{5}{2}$ state does not experience spin reversal and so is suppressed purely as a result of magneto-orbital coupling. In this picture, the $\frac{7}{3}$ gap would presumably increase with a smaller slope than if  spin effects alone were present. The saturation of the $\frac{7}{3}$ gaps at high tilt is evidence for a possible crossover 
taking place between such a spin-dominated and orbitally-coupled regime. 


We emphasize that the naive direct explanation of our data, namely that the $\frac{5}{2}$ $(\frac{7}{3})$ FQH state is spin-unpolarized (polarized) although believed to be unlikely for reasons discussed above, cannot be compellingly ruled out. Since the principal motivation \cite{Morf:PRL:1998}  for accepting the spin-polarized description of the $\frac{5}{2}$ was based on an assumed universal similarity in the behaviour of the $\frac{5}{2}$ and $\frac{7}{3}$ states in a tilted field, our experimental finding that the two states behave qualitatively differently may necessitate a fundamental rethinking of the nature of the $\frac{5}{2}$ FQHE. In this context, it is interesting to mention that some recent theoretical papers \cite{Jain2006,Jain2007} have indeed raised questions regarding the nature of the $\frac{5}{2}$ state, and if further investigations proved it to be non spin-polarized, it would unlikely be a Moore-Read non-Abelian Pfaffian state.


In conclusion, the $\frac{7}{3}$ FQH gap is observed to be enhanced  by an applied parallel magnetic field in contrast to the $\frac{5}{2}$ gap which is strongly suppressed, in spite of the two gaps being energetically comparable at zero parallel fields in our sample. This contrasting dichotomy between these two states is unexpected, and calls into question the prevailing theoretical belief that they should behave similarly if both are spin-polarized. We interpret the behavior of the $\frac{7}{3}$ gap in our experiment to be arising from the formation of skyrmions consisting of a small number of spin-reversed excitations whereas  the suppression of the $\frac{5}{2}$ state in the presence of even a modest parallel field remains an open question.


This work has been supported by NSERC (Canada), the Canadian Institute for Advanced Research, FQRNT (Qu\'ebec), the Alfred P. Sloan Foundation (PH and GG) and the Microsoft Q Project (SDS). We would like to thank H.L. Stormer for helpful discussions, as well as R. Talbot, R. Gagnon and J. Smeros for technical assistance.

\end{document}